\input amstex 
\documentstyle{amsppt} 
\document
\magnification=1000
\baselineskip=14truept
\hsize=5.5in
\vsize=8in
\headline = {\ifodd\pageno Quantum Algorithm Uncertainty Principles \hfil\folio\else\folio\hfil Ken Loo\fi}
\hoffset=.5in
\voffset=.5in
\pageno=215
\baselineskip=12true pt

\topmatter
\title
Quantum Algorithm Uncertainty Principles 
\endtitle

\leftheadtext{Ken Loo}
\rightheadtext{Quantum Algorithm Uncertainty Principles}

\address P.O. Box 9160, Portland, OR. 97207
\endaddress

\email look\@sdf.lonestar.org \endemail



\abstract 
Previously, Bennet and Feynman
asked if Heisenberg's
uncertainty principle puts a limitation on a quantum
computer
(Quantum Mechanical Computers,
Richard P. Feynman, Foundations of Physics,
Vol. 16, No. 6, p597-531,
1986).
Feynman's answer was negative.  In this paper,
we will revisit the same question for the discrete
time Fourier transform uncertainty principle.
We will show that the discrete
time Fourier
transform uncertainty principle plays a fundamental role
in showing that Shor's
type of quantum algorithms has efficient running time
and conclude that
the discrete time uncertainty principle is an aid in
our current formulation and understanding of
Shor's type of quantum algorithms.
It turns out that
for these algorithms,
the probability of measuring an element in 
some set $T$ (at the end of the algorithm) can be
written in terms of the time-limiting and
band-limiting operators from finite Fourier analysis.
Associated with these operators is the finite
Fourier transform uncertainty principle.  The
uncertainty principle provides a lower bound for
the above probability.  We will
derive lower bounds for these types of probabilities in
general.  
We will call these lower bounds
quantum algorithm uncertainty principles or QAUP.  
QAUP are important because they give us some sense
of the probability of measuring something desirable.
We will use these lower bounds to derive Shor's
factoring and discrete log algorithms.   

\endabstract

\author Ken Loo 
\endauthor
\endtopmatter

\subhead{\bf 0.  Introduction}
\endsubhead
Feynman and Bennet (see [2]) previously asked if Heisenberg's
uncertainty principle puts a limitation on quantum
computers.  Feynman's answer was negative.
Given that the quantum Fourier transform is at the core
of Shor's type of quantum algorithms, it is natural to ask
if the finite Fourier transform uncertainty principle
puts limitations on quantum algorithms and quantum computers.
In this
paper, we will show that the discrete time Fourier
transform uncertainty principle is an aid for
Shor's type of quantum algorithms.

It is well known that the uncertainty principle is a property
of Fourier analysis.  Basically, the uncertainty principle
says that a nonzero function $f$ and its Fourier transform
$\hat f$ can not be both highly localized or concentrated.
In applications, the most famous uncertainty principle
is Heisenberg's uncertainty principle.  There are
also less
well known finite Fourier transform uncertainty principles
that are used in signals analysis (see [1] and [6]).   
We will use ideas from
the latter uncertainty principles to derive lower bounds
for the probability of measuring an element in some set $T$
in a certain class of quantum algorithms.  

We will consider quantum algorithms of the
following type.  Start with two registers both
set to $0$.  Apply the quantum Fourier transform
(over $p$)
to the first register.  Compute from the first register
the function $g\left(j\right)$ and store its
value in the second register:
$$
  |0\rangle|0\rangle
  \to\dfrac{1}{\sqrt{p}}
   \sum_{j=0}^{p-1}|j\rangle|0\rangle
  \to\dfrac{1}{\sqrt{p}}
   \sum_{j=0}^{p-1}|j\rangle|g\left(j\right)\rangle . \tag0.1
$$
We then measure the second register and see the value $b$.  
Next, zero-pad the $p$ dimensional quantum state up to a $q$ 
dimensional quantum state for $q \geq p$
and then apply the quantum Fourier transform (over $q$) to the
first register.  Using the notation 
$$
  B = \left\{c\in\Bbb Z_p \bigg| g\left(c\right) = b\right\}, \tag0.2
$$
in mathematical notation, the process is 
$$
  \align
  {}&\dfrac{1}{\sqrt{p}}
   \sum_{j=0}^{p-1}|j\rangle|g\left(j\right)\rangle
   \to\dfrac{1}{\sqrt{|B|}}
   \sum_{c\in B}|c\rangle|b\rangle
   \to\tag0.3\\
   {}&\dfrac{1}{\sqrt{|B|}}
   \dfrac{1}{\sqrt{q}}
   \sum_{k=0}^{q-1} \left\{
   \sum_{c\in B} \exp{\left(
   \dfrac{2i\pi ck}{q}\right)}      
   \right\}|k\rangle|b\rangle \equiv \\
   {}&\dfrac{1}{\sqrt{|B|}}
   \dfrac{1}{\sqrt{q}}
   \sum_{k=0}^{q-1} \left\{
   \sum_{c\in B} \exp{\left(
   \dfrac{2i\pi ck}{q}\right)}
   \right\}|k\rangle ,
\endalign
$$
where we have dropped the second register
in the last expression in 0.3 since 
it no longer concerns us.
Finally, we measure the first register and we would
like to know the probability of measuring an
element in some set $T$ in this last measurement.  
Let us denote this probability by 
$prob\left(T, p, q\right)$.  In particular,
$$
  prob\left(T, p, q\right) = 
   \dfrac{1}{q|B|}
  \sum_{c\in T}
  \Bigg|\sum_{c\in B} \exp{\left(
   \dfrac{2i\pi ck}{q}\right)}
   \Bigg|^2 \tag0.4
$$

In general, the expression in 0.4 is hard to
compute in closed form.  Techniques have been
developed to derive lower bounds for 0.4 and the 
correctness of Shor's factoring and discrete log 
algorithm can be derived this way (see [3]).  
We now rewrite 0.4 in terms of the time-limiting 
and band-limiting operators.  
Let $f = |0\rangle$, then 
$f_j\exp{\left(\frac{-2i\pi jc}{q}\right)}$
is $0$ if $j\neq 0$ and it is $1$ if $j = 0$.  
Using this,  
0.3 can be written as 
$$
  \dfrac{1}{\sqrt{|B|}}
   \dfrac{1}{\sqrt{q}}
   \sum_{k=0}^{q-1} \left\{
   \sum_{c\in B} \left(
    \sum_{j=0}^{q-1}
     f_j\exp{\left(\frac{-2i\pi jc}{q}\right)}
    \right)
    \exp{\left(
   \dfrac{2i\pi ck}{q}\right)}
   \right\}|k\rangle .  \tag0.5
$$
The expression in 0.5 is 
$\sqrt{\frac{q}{|B|}}R_B^qf$ where
$R_B^qf$ is the band-limiting
operator from finite Fourier transform analysis.
In signals analysis language,
the band-limiting operator Fourier
transforms $f$ into the frequency domain,
zeros out the transformed signal outside 
of some band $B$ and then attempts to
reconstruct the original signal $f$ with
an inverse Fourier transform as given
in 0.5.  Also from signals analysis,
for any signal $f$,
the time-limiting operator $P_T^q f$ zeros
out f outside of the set $T$.  With the 
time and band-limiting operators, 0.4
can be written as
$$
  prob\left(T, p, q\right) =
  \dfrac{q}{|B|} ||P_T^q R_B^q f||_2^2 .\tag0.6
$$

The goal of this paper is to derive lower bounds
for 0.6.  This is important since if the
lower bound is of the form
$$
  prob\left(T, p, q\right) >
  \dfrac{1}{poly\left(\text{input size}\right)} , \tag0.7
$$
then by repeating the experiment a polynomial
number of times, we can measure something in 
$T$ with high probability.  We also know that
the quantum Fourier transform can be implemented
with a polynomial number of gates (see [4]).  Hence, given
0.7, we can use a polynomial number of quantum gates
to measure something in $T$ with high probability.  
It turns out that there is
a finite Fourier transform uncertainty principle
associated with the right-hand-side of 0.6.  It is
given by
$$
  \left(1 - \epsilon - \eta\right)||f||_2
  \leq ||P_T^q R_B^q f||_2 \leq 
  ||P_T^q R_B^q || ||f||_2 ,\tag0.8
$$
where $\epsilon$ and $\eta$ depends on $f$.
At first glance, the uncertainty principle
might be useful for quantum computing since
it provides a lower bound for 0.6.  Unfortunately,
0.8 is useless for quantum computing because 
for $f = |0\rangle$, $\epsilon + \eta \geq 1$.
Nonetheless, the proof of the uncertainty
principle in 0.8 was enlightening.  
It paved the way to deriving some general 
lower bounds for 0.6 and these lower bounds
can be used to derive Shor's factoring and
discrete log algorithm.  We will call these lower bounds
quantum algorithm uncertainty principles or QAUP
for short. 
 
It is well known that factoring a positive 
integer $N$ can be reduced to finding
the period $r$ of the function $x^a \bmod N$
for some $x\bmod N$.  Shor (see [4] and [5])
showed us how to do this.
Hallgren and Hales ([3]) abstracted the 
essentials from
Shor's algorithm and derived quantum
Fourier sampling theorems (QFST) to deal
with the type of problems similar to the 
one in the last paragraph.  Basically,
when applied to quantum algorithms,
QFST relate $prob\left(T,p,p\right)$ to
$prob\left(T^{'},p,q\right)$ via equations
of the type
$$
  prob\left(T^{'},p,q\right) \geq
  \dfrac{1}{poly\left(\text{input size}\right)}
  prob\left(T,p,p\right). \tag0.9
$$
Here, $q = poly\left(\text{input size}\right)p$ and
$T^{'}$ is nicely related
to $T$ (for the factoring problem,
$T^{'} = \left\{\lfloor\frac{qj}{p}\rfloor
\Bigg| j\in T\right\}$ ).   Thus, if
$$
  prob\left(T,p,p\right) >
  \dfrac{1}{poly\left(\text{input size}\right)},
  \tag0.10
$$
then
$$
  prob\left(T^{'},p,q\right) >
  \dfrac{1}{poly\left(\text{input size}\right)}.
  \tag0.11
$$
This tells us that we can apply the 
quantum Fourier transform
over some smooth $q > p$ and we are in business
to measure something in $T^{'}$.  

The proof of the uncertainty principle that we are interested
in compares $f$ to $P_T^q R_B^q f$.  The theorems derived
in [3] compares $prob\left(T^{'},p,q\right)$ to
$prob\left(T,p,p\right)$.  It turns out that the proof
of the uncertainity principle can be modified to compare
$prob\left(T^{'},p,q\right)$ to
$prob\left(T,p,p\right)$ also.
Let us write
$$
 prob\left(T,p,p\right) = 
 \sum_{k\in T} prob\left(k,p,p\right).  \tag0.12
$$
The type of lower bounds that we will derive is of 
the form
$$
    prob\left(T^{'},p,q\right) \geq
    \dfrac{p}{q}
    \sum_{k\in T}
    \left(\sqrt{prob\left(k,p,p\right)} -
     small\left(k,\dots\right) \right)^2 ,
    \tag0.13
$$
for $q > p$.  Here, $small\left(k, \dots\right)$
is some small number that depends on all the
variables involved, $T$ and $T^{'}$ do not have
to be nicely related to each other as long as 
for all values of $k$, the expression that gets 
squared in 0.13 is positive.  

After deriving quantum algorithm uncertainty
principles, we will apply them to the
factoring and discrete log problems.  
For factoring,
$\sqrt{prob\left(k,p,p\right)} = \frac{1}{\sqrt{r}}$
for all $k$ and $|T| = \Phi\left(r\right)$, where
$\Phi\left(r\right)$ is the Euler phi function.  
After some computation, 
$\sqrt{prob\left(k,p,p\right)} = 
\frac{1}{\sqrt{r}}$ factors
out of the right-hand-side
of 0.13 leaving
$$\align
  {}&prob\left(T^{'},p,q\right) \geq 
  \dfrac{p}{q}\left(1 - small\right)^2
  \sum_{c\in T}\dfrac{1}{r} =
  \dfrac{p}{q}\left(1 - small\right)^2
  \dfrac{\Phi\left(r\right)}{r} > \tag0.14 \\
  {}&\dfrac{p}{q}\left(1 - small\right)^2
  \dfrac{1}{\log\log r} >
  \dfrac{p}{q}\left(1 - small\right)^2
  \dfrac{1}{poly\left(\log r\right)}. 
\endalign
$$
This tells us that it is OK to transform
over some smooth $q > p = rt$.  As in Shor's
algorithm (see [4]), after measuring
something in $T^{'}$, $r$ can be 
recovered by using continued fractions.    
The discrete log problem is similar.

It is natural to ask if quantum algorithm
uncertainty principles have any physical
significance other than as a tool for 
proving theorems.  This is not an easy
question to answer since  
we currently do not have
the concepts of time and
frequency domain in quantum
computing.  One might venture
to ask if the power of quantum computing
is somehow related to the uncertainty
principle (Heisenberg or QAUP).  If
it were, it would 
certainly be a triumph for the uncertainty 
principle.  We will leave these 
questions open for future research.  
We will end our introduction with a quote
from [1].  ``The uncertainty
principle is widely known for its
philosophical applications: in quantum
mechanics, of course, it shows that a particle's 
position and momentum can not be determined
simultaneously...; in signal processing it 
establishes limits on the extent to
which the instantaneous frequency of
a signal can be measured... However,
it also has technical applications, for example
in the theory of partial differential equations..."
- D. Donoho and P. Stark    

\subhead{\bf 1. Finite Fourier Transform Uncertainty Principles}
\endsubhead
We start by giving a review of the finite 
Fourier transform uncertainty principles.
The notations in this section are taken from [6] and 
they are slightly different
from standard quantum computing notations.  
Let $f:\Bbb Z_q\to\Bbb C$,
then the Fourier transform of $f$ is the function given by
$$
  \hat f\left(y\right) = \sum_{x = 0}^{q - 1} f\left(x\right)
   \exp{\left(\frac{-2\pi i x y}{q}\right)} . \tag1.1
$$
The inverse Fourier transform of $\hat f$ is the function given by
$$
   f\left(x\right) = \dfrac{1}{q}\sum_{y = 0}^{q - 1} \hat f\left(y\right)
   \exp{\left(\frac{2\pi i x y}{q}\right)} . \tag1.2
$$
Notice that the above discrete Fourier transform is not the
same transform as the quantum Fourier transform given in equation 3.1,
they differ by a minus sign in the phase of the transform and an
overall normalization constant.  We will use a hat to denote the
discrete Fourier transform and $\text{FT}_p$ to denote the quantum
Fourier transform over $p$.

In vector or quantum computing notation, $f$ is a vector given by
$$
  f = \sum_{j = 0}^{q - 1} f_j |j\rangle \tag1.3
$$
where $f_j = f\left(j\right)$.  The Fourier and inverse Fourier transform are given
by
$$
  \hat f = \sum_{j = 0}^{q - 1} \left\{\sum_{l = 0}^{q - 1} f_l
   \exp{\left(\frac{-2\pi i l j}{q}\right)}\right\} |j\rangle ,\tag1.4
$$
and
$$
  f = \dfrac{1}{q}\sum_{j = 0}^{q - 1} \left\{\sum_{l = 0}^{q - 1} \hat f_l
   \exp{\left(\frac{2\pi i l j}{q}\right)}\right\} |j\rangle ,\tag1.5
$$
respectively.
\proclaim{Definition 1.1} The support of $f$ is defined to be
$$
  \text{supp} f = \left\{x\in \Bbb Z_q | f\left(x\right) \neq 0\right\} .
   \tag1.6
$$
\endproclaim

\proclaim{Theorem 1.1 \bf Uncertainty Principle - version 1} Suppose
$f:\Bbb Z_q\to \Bbb C$ and $f$ is not the zero function, then
$|\text{supp }f||\text{supp }\hat f| \geq q$.
\endproclaim
\demo{Proof} See [6] and references within.  \qed
\enddemo

\proclaim{Definition 1.2} Let
$T\subseteq \Bbb Z_q$,  define $\delta_T:\Bbb Z_q\to \left\{0, 1\right\}$ by
$$
     \delta_T\left(x\right) =
     \cases 1, & \text{if } x\in T \\
  0, & \text{otherwise}.
\endcases \tag1.7
$$
\endproclaim

\proclaim{Definition 1.3 Time and Band-limiting Operators}
Let $f:\Bbb Z_q\to \Bbb C$ and $B, T\subseteq \Bbb Z_q$, the time-limiting and
band-limiting operators are defined by
$$
  \left[P_T^q f\right]\left(x\right) = 
  f\left(x\right)\delta_T\left(x\right), \tag1.8
$$
and
$$\align
 \left[R_B^qf\right]\left(x\right) {}& = 
   \dfrac{1}{q}\sum_{c \in B} \hat f\left(c\right)
   \exp{\left(\frac{2\pi i x c}{q}\right)}\tag1.9 \\
    {}& = \dfrac{1}{q}\sum_{c \in B}
     \left\{\sum_{j = 0}^{q - 1} f\left(j\right)
   \exp{\left(\frac{-2\pi i j c}{q}\right)}\right\}
   \exp{\left(\frac{2\pi i x c}{q}\right)}, 
\endalign
$$
respectively.
\endproclaim
The time-limiting operator zeros out $f$ outside of 
$T$ and the band-limiting operator
attempts to reconstruct $f$ from $\hat f$ by using 
only information within the set (band) $B$.
In vector notation, the band-limiting operator is given by
$$
  R_B^q f = \dfrac{1}{q} \sum_{k=0}^{q-1}\left\{\sum_{c \in B}
     \left\{\sum_{j = 0}^{q - 1} f_j
   \exp{\left(\frac{-2\pi i j c}{q}\right)}\right\}
   \exp{\left(\frac{2\pi i k c}{q}\right)}\right\}|k\rangle . \tag1.10
$$
From here on, we will assume that $T\neq \emptyset$ 
and $B\neq\emptyset$ since the empty set is not
very interesting to us.

The 2-norm of $f:\Bbb Z_q \to \Bbb C$ is defined by
$$
  ||f||_2 = \sqrt{\langle f, f \rangle} =
   \sqrt{\sum_{x = 0}^{q - 1} |f\left(x\right)|^2 } . \tag1.11
$$
It follows from Parseval's equality that 
$\langle f, f\rangle = \frac{1}{q} \langle \hat f, \hat f\rangle$
and 1.11 implies that 
$||f||_2 = \frac{1}{\sqrt{q}}||\hat f||_2$.
Let $F = \left\{f:\Bbb Z_q \to \Bbb C\right\}$ and $Q$ be a linear operator
$Q:F\to F$.  The operator norm of $Q$ is defined by
$$
  ||Q|| = \sup_{f\neq 0}\left\{\dfrac{||Qf||_2}{||f||_2}\right\}.
  \tag1.12
$$
\proclaim{Lemma 1.1} Let $T, B\subseteq \Bbb Z_q$, then
$||R_B^q|| = 1, ||P_T^q|| = 1$, and
$$
  ||P_T^q R_B^q|| = ||R_B^q P_T^q|| \leq 1 .
  \tag1.13
$$
\endproclaim
\demo{Proof} See [6]. \qed
\enddemo

\proclaim{Theorem 1.2 \bf Uncertainty Principle - version 2}
$$
  \dfrac{\sqrt{|T||B|}}{q} \leq ||P_T^q R_B^q|| \leq
   \sqrt{\dfrac{|T||B|}{q}}.
   \tag1.14
$$
\endproclaim
\demo{Proof} See [6] . \qed
\enddemo
\proclaim{Definition 1.4} We say that $f:\Bbb Z_q\to\Bbb C$ is
$\epsilon$-concentrated on $T\subseteq \Bbb Z_q$ if
$$
  ||f - \delta_T f||_2 = ||f - P_T^q f||_2 \leq \epsilon ||f||_2 ,
  \tag1.15
$$
and we say that $f$ is $\eta$-band-limited to $B$ if there is a function
$f_B:\Bbb Z_q\to \Bbb C$ such that $\text{supp}(\hat f_B) \subseteq B$
and
$$
  ||f - f_B||_2 = \dfrac{1}{\sqrt{q}}||\hat f - \hat f_B||_2 \leq \eta ||f||_2 .  \tag1.16
$$
\endproclaim
We now present the last finite Fourier transform uncertainty principle.
This uncertainty principle is the one that we are interested in for
applications to quantum algorithms.
We will give a proof of this version of the uncertainty
principle since it compares $f$ to 
$P_T^q R_B^q f$ as mentioned earlier.  We will 
modify this theorem to compare 
$prob\left(T^{'},p,q\right)$ to
$prob\left(T,p,p\right)$.
\proclaim{Theorem 1.3 \bf Uncertainty Principle - version 3}
Suppose $f:\Bbb Z_q\to \Bbb C$ is nonzero and it is
$\epsilon$-concentrated on $T$ as well as $\eta$-band-limited on $B$,
then
$$
 \left(1 - \epsilon - \eta\right)||f||_2 \leq ||P_T^q R_B^q f||_2,
 \tag1.17
$$
and
$$
  1 - \epsilon - \eta \leq ||P_T^q R_B^q|| \leq \sqrt{\dfrac{|T||B|}{q}} .
  \tag1.18
$$
\endproclaim
\demo{Proof (From [6]) } We have
$$
  \align
  ||f||_2 - ||P_T^q R_B^q f||_2 {}& \leq
  ||f - P_T^q R_B^q f||_2 \tag1.19 \\
   {}& \leq
  ||f - P_T^q f||_2 + ||P_T^q f - P_T^q R_B^q f||_2 \\
   {}& \leq \epsilon ||f||_2 + ||P_T^q|| ||f - R_B^q f||_2 \\ 
   {}& \leq \epsilon ||f||_2 + ||f - R_B^q f||_2 ,
\endalign
$$
where we have used
$||P_T^q|| = 1$ from lemma 1.1.
Since $f$ is $\eta$-band-limited on $B$, it follows that
$$\align
  ||f - R_B^q f||_2 {}& =
  \dfrac{1}{\sqrt{q}}||\hat f - P_B^q\hat f||_2  \tag1.20\\
  {}& \leq \dfrac{1}{\sqrt{q}}||\hat f - f_B||_2 \leq \eta ||f||_2 .
\endalign
$$
This gives
$$
  ||f||_2 - ||P_T^q R_B^q f||_2 \leq
   \epsilon ||f||_2 + \eta ||f||_2 ,\tag1.21
$$
and $\left(1 - \epsilon - \eta\right)||f||_2 \leq ||P_T^q R_B^q f||_2$ follows.
Finally, $\left(1 - \epsilon - \eta\right) \leq ||P_T^q R_B^q||$ follows from
$||P_T^q R_B^q f||_2 \leq \left(||P_T^q R_B^q||\right) ||f||_2$, and
$||P_T^q R_B^q|| \leq \sqrt{\dfrac{|T||B|}{q}}$
is from uncertainty principle-version 2. \qed
\enddemo

\subhead{\bf 2. Uncertainty in Quantum Algorithms} 
\endsubhead
We now show how uncertainty
principle - version 3 can be
applied to quantum algorithms.
We will consider     
quantum algorithms of the following form.  Prepare $|0\rangle |0\rangle$.  
Apply the quantum
Fourier transform (eqn. 3.1) 
to the prepared state
$$
  |0\rangle |0\rangle\to
  \dfrac{1}{\sqrt{p}}\sum_{j = 0}^{p-1} |j\rangle |0\rangle .
   \tag2.1
$$  
Next, compute
$$
  \dfrac{1}{\sqrt{p}}\sum_{j = 0}^{p-1} |j\rangle |0\rangle \to
  \dfrac{1}{\sqrt{p}}
  \sum_{j = 0}^{p-1} |j\rangle |g\left(j\right)\rangle 
  \tag2.2
$$ 
and then measure the second register 
obtaining $b = g\left(c\right)$.  This will
put the quantum computer into the state 
$$
   \dfrac{1}{\sqrt{|B|}}\sum_{c\in B}|c\rangle ,
   \tag2.3
$$ 
where $B = \left\{c| c\in\Bbb Z_p, b = g\left(c\right)\right\}$ 
and the second register is suppressed.  
We then zero-pad the quantum state from
a $p$-dimensional state to a $q$-dimensional state
for some $q \geq p$ and apply the quantum Fourier transform over 
$\Bbb Z_q$ to the first register.  The 
quantum computer will be in the state
$$
  \dfrac{1}{\sqrt{|B|}}
  \dfrac{1}{\sqrt{q}}
  \sum_{k = 0}^{q - 1}\left\{
   \sum_{c\in B}\exp{\frac{2i\pi ck}{q}} \right\}
  |k\rangle . \tag2.4
$$
Finally, we measure the first register and the probability 
of measuring an element in some set 
$T \subseteq \Bbb Z_q$ is of great interest.  
In particular, we would like this probability to be 
at least  
$\frac{1}{poly\left(\text{input size}\right)}$ so 
that the experiment can be repeated
a polynomial number of times in hope of measuring something in 
$T$ with high probability.  

It turns out that the probability of measuring something in $T$ can be 
written in terms of the time-limiting and band-limiting operators from
section 1.  Let $f = |0\rangle$, then 2.4 
can be written as 
$$
  \dfrac{1}{\sqrt{|B|}}
  \dfrac{1}{\sqrt{q}}
  \sum_{k = 0}^{q - 1}\left\{
   \sum_{c\in B}
    \left(
    \sum_{j=0}^{q - 1} f_j 
     \exp{\left(\frac{-2i\pi jc}{q}\right)}
     \right)
    \exp{\frac{2i\pi ck}{q}} \right\}
  |k\rangle  = 
   \frac{\sqrt{q}}{\sqrt{|B|}} R_B^q f  , 
  \tag2.5
$$
where the equality comes from 1.10.
The probability of measuring something in
the set $T$ is just 
$$
  \frac{q}{|B|} ||P_T^qR_B^q f||_2^2. \tag2.6
$$  
 
At first glance, the uncertainty principle - version 3 might provide
a useful lower bound for the probability of measuring something in
$T$.  It turns out that version 3 is totally useless for this purpose.
If $0\notin T$, $f$ is 1-concentrated on $T$ ($\epsilon$ is at least 1)
and the uncertainty inequality is rendered useless.  If $0\in T$, then
$\epsilon = 0$ and we take $\eta = \sqrt{\frac{q - |B|}{q}}$ 
(it is at least that big).  The probability of measuring $0$ in the
first register is $\frac{|B|}{q}$, and the uncertainty principle
tells us that the probability of measuring
something in $T - \left\{0\right\}$ is greater than
$$
  \dfrac{q}{|B|}\left(1 - \sqrt{\dfrac{q - |B|}{q}}\right)^2  
  - \dfrac{|B|}{q} . \tag2.7
$$
The expression in 2.7 is negative!

Nonetheless, the proof of theorem 1.3 is a good model for 
deriving useful lower bounds.
In light of theorem 1.3, and the work of [3], 
it turns out that it is fruitful 
to derive uncertainty principles which relate 
$||P_{T^{'}}^qR_B^q f||_2$
to $||P_T^pR_B^p f||_2$ for $q > p$.  We will do just that. 

\subhead{\bf 3. Motivations: Factoring, The Easy Case} 
\endsubhead
In this section, we will review an easy case of
Shor's factoring algorithm.
Factoring an odd integer $N$ can be reduced to finding 
the order (even) of a random integer $x\bmod N$, i.e. 
finding $r$ such that $x^r\equiv 1\bmod N$ where $r$ is divisible by 2.  
Suppose $N$ has $k$ distinct prime factors, i.e.
$N = p_1^{e_1}\dots p_k^{e_k}$,
then algorithm is as follows:  

Step 1. Choose a random $x \bmod N$.  Use the polynomial time Euclidean algorithm to compute gcd$(x, N)$. 
If gcd$(x, N) \neq 1$, a factor of N is found so assume gcd$(x, N) = 1$.

Step 2. Find the smallest $r$ such that $x^r \equiv 1 \bmod N$.  This is the "hard" part of the algorithm.

Step 3. If $r$ is odd, go back to step 1.  The probability that $r$ is odd is 
$\left(\frac{1}{2}\right)^k$.  

Step 4. Since $r$ is even, 
$\left(x^{\frac{r}{2}} - 1\right)\left(x^{\frac{r}{2}} + 1\right) =
  x^r - 1 \equiv 0 \bmod N$.  
If $\left(x^{\frac{r}{2}} + 1\right)\equiv 0 \bmod N$, go back to step 1.
The probability that $\left(x^{\frac{r}{2}} + 1\right)\equiv 0 \bmod N$
is less than $\left(\frac{1}{2}\right)^{k-1}$.

Step 5. Since $\left(x^{\frac{r}{2}} + 1\right)\neq 0 \bmod N$, 
$\left(x^{\frac{r}{2}} - 1\right)$ and $N$ must have a nontrivial 
common factor.  Use the polynomial time Euclidean algorithm to
find gcd$(x^{\frac{r}{2}} - 1, N)$.  
Notice that gcd$(x^{\frac{r}{2}} - 1, N) \neq N$ otherwise
$ x^{\frac{r}{2}} \equiv 1 \bmod N$ contradicting step 2 where
$r$ is the smallest such number with this property.

The hard part of the above algorithm is in step 2, but 
Shor [4] and [5] showed us how to do that step with a polynomial 
number of quantum gates.  His algorithm relies heavily on 
the quantum Fourier transform.  
The quantum Fourier transform is a unitary 
operator that takes the state 
$|a\rangle = \sum_{j=0}^{p-1} a_j|j\rangle$ to
$$
  \text{FT}_p |a\rangle = 
  \dfrac{1}{\sqrt{p}}\sum_{c = 0}^{p-1} 
   \left[\sum_{j = 0}^{p-1} a_j\exp{\left(2\pi ijc/p\right)}
    \right]|c\rangle,  
   \tag3.1
$$
where $p = 2^l$ ($p$ can be a smooth integer). 
The quantum Fourier transform can be implemented with a polynomial 
number of quantum gates, see [4].  
We will review Shor's algorithm for a simple 
(rather unrealistic) case.

We start with $|0\rangle |0\rangle$ and take the Fourier transform of the first register 
$$
  |0\rangle |0\rangle \to 
  \dfrac{1}{\sqrt{p}}\sum_{c = 0}^{p-1} |c\rangle |0\rangle  .
  \tag3.2
$$
Next, we perform the operation
$$ 
  \dfrac{1}{\sqrt{p}}\sum_{c = 0}^{p-1} 
  |c\rangle |0\rangle \to \dfrac{1}{\sqrt{p}}
  \sum_{c = 0}^{p-1} |c\rangle |x^c\bmod N\rangle .
  \tag3.3
$$
Here is where we make our unreasonable assumption.  
Assume that the period $r$ exactly divides $p$ and write $p = rt$.  
Now we measure the second register and see $b = x^a\bmod N$,  
where a is the smallest positive integer that satisfies the equation.  
Notice that we have no knowledge of $a$.  
This puts our quantum computer into the state
$$
  \dfrac{1}{\sqrt{A + 1}}\sum_{j=0}^A |j*r + a\rangle |b\rangle
  \tag3.4
$$
where $A$ is the largest integer for which $Ar + a < p$.  Since 
$p=rt$, $A = \frac{p}{r} - 1$. The above state can be rewritten as
$$ 
  \sqrt{\dfrac{r}{p}}\sum_{j=0}^{\dfrac{p}{r} - 1}|j*r + a\rangle, 
   \tag3.5
$$
where the state $|b\rangle$ has been dropped for notation convenience.
Next, apply the quantum Fourier transform:
$$\align
  \sqrt{\dfrac{r}{p}}
   \sum_{j=0}^{\dfrac{p}{r} - 1}
   |j*r + a\rangle {}& \to 
   \dfrac{1}{\sqrt{p}}
   \sum_{c = 0}^{p-1} 
   \sqrt{\dfrac{r}{p}}
   \sum_{j=0}^{\dfrac{p}{r} - 1} 
   \exp{\left(2\pi i(jr + a)c/p\right)} |c\rangle \tag3.6\\
  {}& = \dfrac{\sqrt{r}}{p}
  \sum_{c = 0}^{p-1} 
   \exp{\left(2\pi iac/p\right)} 
    \sum_{j=0}^{\dfrac{p}{r} - 1} 
    \exp{\left(2\pi ijrc/p\right)} |c\rangle ,
\endalign
$$
      
Observing that 
$$
  \sum_{j=0}^{\dfrac{p}{r} - 1} \exp{\left(2\pi ijrc/p\right)} =
  \cases \dfrac{p}{r}, & \text{if }c\text{ is a multiple of } \dfrac{p}{r} \\
  0, & \text{otherwise}     
\endcases  ,
  \tag3.7
$$
eqn. 3.6 is just 
$$
  \dfrac{1}{\sqrt{r}}
  \sum_{j=0}^{r-1} 
  \exp{\left(2\pi iaj/r\right)} 
   |jp/r\rangle  = 
  \dfrac{1}{\sqrt{r}}
  \sum_{j=0}^{r-1} 
  \exp{\left(2\pi iaj/r\right)} |jt\rangle .
  \tag3.8
$$
Finally, with probability $\frac{1}{r}$,
we measure the first register and see $k = \dfrac{jp}{r} = jt$ where 
$0 \leq j \leq r-1$.  
Notice that the probability of measuring 
$|k\rangle|b\rangle$ for any particular $k$ and $b$ is
$1/r^2$.
This gives us $\dfrac{k}{p} = \dfrac{j}{r}$ where $k$ and $p$ are known.  
If $j$ and $r$ are relatively prime, we can determine $r$.  
The number of $j$'s that are relatively prime to $r$ 
is $\bold\Phi\left(r\right)$.
Since there are $r$ number of different $k$'s and $b$'s, 
the total probability of observing 
$|k=jt\rangle |b\rangle$ such that $j$ is relatively prime to $r$ is 
$$
   r^2\dfrac{\bold\Phi\left(r\right)}{r}\dfrac{1}{r^2} = 
    \dfrac{\bold\Phi\left(r\right)}{r} > \dfrac{k}{\log\log r} \equiv
   \dfrac{k}{\log n} . 
   \tag3.9
$$ 
This tells us that for the simple case, with high probability, 
we can find $r$ with a polynomial 
(with respect to $n = \log r$) number of quantum gates.
In other words, let 
$$
  T = \left\{k = jt |\gcd\left(j, r\right) = 1,
  0 \leq j \leq r - 1\right\}.
  \tag3.10
$$
Then we can measure something in $T$ with probability
at least the inverse of a polynomial in the
size of the input. 

The above algorithm is not very useful
in general.  The reason for this is that
we assumed $p = rt$ and we do not
know $r$ ($r$ is what we seek).  
We performed two quantum
Fourier transforms over $p$.  
In generalizing this easy case to the general
case, we must address these two
Fourier transforms.

\subhead{\bf 4. Quantum Algorithm Uncertainty Principles (QAUP) - version 1} 
\endsubhead
We now derive quantum algorithm uncertainty principles that relate
$||P_{T^{'}}^qR_B^q f||_2$
to $||P_T^pR_B^p f||_2$ for $q > p$.  

\proclaim{Proposition 4.1} For all $x\in\Bbb R$,
$|\exp{\left(ix\right)} - 1| \leq |x|$.
\endproclaim
\demo{Proof} The inequality is certainly true when $2 \leq |x|$, 
so we will assume $2 > |x|$.
The expression $|\exp{\left(ix\right)} - 1|$ is the length 
of the vector $\left(\cos x, \sin x\right) - (1, 0)$ in $\Bbb R^2$.  
The smallest arc-length of the piece of the unit
circle that starts at $\left(0,1\right)$ and ends at
$\left(\cos x, \sin x\right)$ is $1*|x|$.  Since the
shortest distance between two points in $R^2$ is a straight line,
$|\exp{\left(ix\right)} - 1| \leq |x|$ follows.
\enddemo

\proclaim{Lemma 4.1} Let $q > p$,  $f = |0\rangle$, 
$T\subseteq\Bbb Z_p$, and  
$$ 
    T^{'} = \left\{k^{'}\in\Bbb Z_{q}\Bigg| k^{'} = \dfrac{qk}{p} + \epsilon_k,
    k\in T, 
   \text{ no restrictions on } \epsilon_k 
\right\} . \tag4.1
$$
Let $\bar B$ be a set and 
$B = a + \bar B = \left\{a + b|b\in\bar B\right\}$.
Suppose that for all $k\in T$, 
$$
  0  
    \leq 
   \dfrac{2\pi |\epsilon_k|}{q^2}
    \sum_{c\in\bar B} |c| \leq 
   \dfrac{2\pi \delta_{k,\bar B}}{q^2}
    \leq 
    \dfrac{p}{q} ||P_k^pR_B^p f||_2 ,
    \tag4.2
$$
where $\delta_{k,\bar B}$ depends on 
$k$ and $\sum_{c\in \bar B}|c|$.
Then for all $k^{'}\in T$, 
$$\align
  {}& 0 \leq 
  \dfrac{p}{q} ||P_k^pR_B^p f||_2 -
   \dfrac{2\pi \delta_{k,\bar B}}{q^2} \leq \tag4.3 \\
  {}&\dfrac{p}{q} ||P_k^pR_B^p f||_2 -
   \dfrac{2\pi |\epsilon_k|}{q^2}
    \sum_{c\in \bar B} |c| 
    \leq \\
    {}& ||P_{k^{'}}^qR_B^q f||_2 .
\endalign
$$
\endproclaim
\demo{Proof} The left most two inequalities 
in 4.3 comes from 4.2.  Notice
that the relationship between $B$, $\bar B$,
0.4, and 2.6 imply that
$$\align
  {}& ||P_k^pR_B^p f||_2 = \dfrac{1}{p}
      \Bigg|\sum_{c\in B}\exp{\left(\dfrac{2i\pi ck}{p}\right)}\Bigg| 
        = \dfrac{1}{p}
      \Bigg|\sum_{\bar c\in\bar B}
      \exp{\left(\dfrac{2i\pi\bar ck}{p}\right)}\Bigg|, 
      \tag4.4 \\
  {}& ||P_{k^{'}}^qR_B^q f||_2 = \dfrac{1}{q}
      \Bigg|\sum_{c\in B}\exp{\left(\dfrac{2i\pi ck^{'}}{q}\right)}\Bigg| 
        = \dfrac{1}{q}
      \Bigg|\sum_{\bar c\in\bar B}
       \exp{\left(\dfrac{2i\pi\bar ck^{'}}{q}\right)}\Bigg| ,
      \\
  {}& prob\left(k, p, p\right) = 
      \dfrac{p}{|B|}||P_k^pR_B^p f||_2^2 = \dfrac{1}{p|B|}
      \Bigg|\sum_{\bar c\in\bar B}
      \exp{\left(\dfrac{2i\pi\bar ck}{p}\right)}\Bigg|^2, 
       \\
  {}& prob\left(k^{'}, p, q\right) = 
      \dfrac{q}{|B|}||P_{k^{'}}^qR_B^q f||_2^2 = \dfrac{1}{q|B|}
      \Bigg|\sum_{\bar c\in\bar B}
      \exp{\left(\dfrac{2i\pi\bar ck^{'}}{q}\right)}\Bigg|^2, 
\endalign
$$
which imply
$$\align
  {}& \dfrac{p}{q} ||P_k^pR_B^p f||_2 -
      ||P_{k^{'}}^qR_B^q f||_2 \leq
      \Bigg|\dfrac{1}{q} 
      \sum_{\bar c\in\bar B}
      \exp{\left(\dfrac{2i\pi\bar ck}{p}\right)} -
       \dfrac{1}{q}
      \sum_{\bar c\in\bar B}
      \exp{\left(\dfrac{2i\pi\bar ck^{'}}{q}\right)}\Bigg|.
      \tag4.5
\endalign
$$
The expression on the right can be rewritten as 
$$\align
  {}&\dfrac{1}{q}\Bigg|\sum_{\bar c\in\bar B}
      \exp{\left(\dfrac{2i\pi\bar ck}{p}\right)}
       \left(\exp{\left(\dfrac{2i\pi\bar c\epsilon_k}{q}\right)}
        - 1 
       \right)\Bigg| . \tag4.6
\endalign
$$  
Using 
$$
  |\exp{\left(ix\right)} - 1| < |x|, \tag4.7
$$
4.6 can be bounded from above by
$$\align
  {}&\dfrac{1}{q}\sum_{\bar c\in\bar B}
      \Bigg|
           \left(\exp{\left(\dfrac{2i\pi\bar c\epsilon_k}{q}\right)}
        - 1
       \right)\Bigg| \leq
     \dfrac{1}{q}\sum_{\bar c\in\bar B}
           \Bigg|\dfrac{2i\pi\bar c\epsilon_k}{q}\Bigg|
      = \tag4.8 \\
      {}& \dfrac{2\pi |\epsilon_k|}{q^2}
      \sum_{\bar c\in \bar B} |\bar c| \leq 
       \dfrac{2\pi \delta_{k,\bar B}}{q^2} .
\endalign
$$
Finally, 4.5, 4.6 and 4.8 imply 4.3 \qed
\enddemo

\proclaim{Theorem 4.1 QAUP - version 1}
Under the conditions of lemma 4.1, 
$$  
  0 \leq \dfrac{p}{q}
   \left\{  
     \dfrac{\sqrt{p}}{\sqrt{|B|}} 
   ||P_k^pR_B^p f||_2 -
   \dfrac{2\pi \delta_{k,\bar B}}{q\sqrt{|B|p} }
   \right\}^2  \leq
      \dfrac{q}{|B|}
      ||P_{k^{'}}^qR_B^q f||_2^2 . \tag4.9
$$ 
\endproclaim

\demo{Proof} Since 4.3 is greater than zero,
we can square all terms in 4.3 and keep
the inequalities and obtain
$$ 
  \align
  0 {}&\leq 
  \dfrac{q}{|B|}
  \left\{
   \dfrac{\sqrt{p|B|}}{q}
     \left(\dfrac{\sqrt{p}}{\sqrt{|B|}}
   ||P_k^pR_B^p f||\right) -
   \dfrac{2\pi\delta_{k,\bar B}}{q^2} 
    \right\}^2
     \tag4.10 \\
     {}& \leq 
      \dfrac{q}{|B|}
      ||P_{k^{'}}^qR_B^q f||^2 . 
\endalign
$$
The middle term in 4.10 is 
$$ 
  \dfrac{p}{q}
   \left\{ 
     \dfrac{\sqrt{p}}{\sqrt{|B|}}
   ||P_k^pR_B^p f||_2 -
   \dfrac{2\pi \delta_{k,\bar B}}{q\sqrt{|B|p} }
   \right\}^2 . \qed \tag4.11
$$
\enddemo
\proclaim{Corollary 4.1 QAUP - version 1.a} 
Under the conditions of lemma 4.1,
let
$$
    T^{'} = \left\{k^{'}\in\Bbb Z_q\Bigg| k^{'} = 
     \lfloor\dfrac{qk}{p}\rfloor ,
      k\in T
     \right\} , \tag4.12
$$
or
$$
    T^{'} = \left\{k^{'}\in\Bbb Z_q\Bigg| k^{'} = 
     \lfloor\dfrac{qk}{p}\rceil , 
      k\in T 
     \right\} ,  \tag4.13
$$ 
or
$$ 
    T^{'} = \left\{k^{'}\in\Bbb Z_q\Bigg| k^{'} = 
     \lceil\dfrac{qk}{p}\rceil , 
      k\in T  
     \right\} .  \tag4.14
$$  
Suppose, 
$$
    \sum_{\bar c\in\bar B} |\bar c|  \leq 
    \delta_{\bar B}, \tag4.15
$$
then
$$
  0 \leq \dfrac{p}{q}
   \left\{
     \dfrac{\sqrt{p}}{\sqrt{|B|}}
   ||P_k^pR_B^p f||_2 -
   \dfrac{2\pi\delta_{\bar B} }{q\sqrt{|B|p} }
   \right\}^2  \leq
      \dfrac{q}{|B|}
      ||P_{k^{'}}^qR_B^q f||_2^2 . \tag4.16
$$
\endproclaim
\demo{Proof} This is an application of $|\epsilon_k| < 1$ 
in 4.2 to 4.9. \qed
\enddemo
\remark{Remark 4.1}  Summing over all $k^{'}\in T$ and
$k\in T$ in 4.16, 4.9 or 4.3 gives inequalities of the
form 0.13.
\endremark

\subhead{\bf 5. Application to Factoring, QAUP - version 1}
\endsubhead
We will now 
show how corollary 4.1 can be applied to factoring.  
Recall that for easy factoring,
$prob\left(k,p,p\right) = \frac{1}{r}$ for any $k$ in
3.10.

Let $T$ be as given in 3.10 and $T^{'}$ be as given in 4.12.
From section 3, we have $B = \left\{a, a + r, \dots , 
a + \left(t-1\right)r \right\}$
which implies 
$|B| = t$ and 
$\bar B = \left\{0, r, 2r, \dots ,\left(t - 1\right)r\right\}$. 
Let us take $p = rt \geq 2r^2$ (i.e. $t \geq 2r$), $q = sp = 2^l > p$
for some $s$ such that
$$
   0 < 1 - \frac{\pi}{s} .
   \tag5.1
$$
The sum $\sum_{\bar c\in\bar B} \bar c$ can
be evaluated in closed form, but it is more convenient to 
bound it from above by an integral and obtain
$$
  \sum_{\bar c\in\bar B} \bar c < r\dfrac{t^2}{2} 
     = \delta_{B}. \tag5.2
$$
With 5.2, 4.4, and our choice of $p$ and $q$,
$$
  \dfrac{2\pi\delta_{B}}{q^2} =
  \dfrac{2\pi}{s^2r^2t^2}\dfrac{rt^2}{2} = 
  \dfrac{\pi}{s^2r}, \tag5.3
$$
and
$$
  \dfrac{p}{q} ||P_k^pR_B^p f||_2 =
  \dfrac{p}{q}\sqrt{\dfrac{|B|}{p}}
   \sqrt{prob\left(k, p, p\right)} =
   \dfrac{1}{s}\sqrt{\dfrac{t}{rt}}
    \dfrac{1}{\sqrt{r}} = 
    \dfrac{1}{sr}. \tag5.4
$$
The expressions in 5.1, 5.3 and 5.4 imply
that 4.2 is satisfied.  

We now proceed to evaluate 4.16 giving
$$
  \dfrac{1}{s}\left\{
  \dfrac{1}{\sqrt{r}} -
   \dfrac{2\pi}{srt\sqrt{t}\sqrt{rt}}
   \dfrac{rt^2}{2}\right\}^2 =
   \dfrac{1}{sr}
   \left(1 - \dfrac{\pi}{s}\right)^2 <
   prob\left(k^{'}, p, q\right). \tag5.5
$$
Finally, we arrive at the desired result.
From 4.16 and 5.5, we have
$$
     prob\left(T^{'}, p, q\right) = 
     \sum_{k^{'}\in T^{'}} \dfrac{q}{|B|}
      ||P_{k^{'}}^qR_B^q f||_2^2 >
     \sum_{k\in T} 
   \dfrac{1}{sr}
   \left(1 - \dfrac{\pi}{s}\right)^2 =
   \left(1 - \dfrac{\pi}{s}\right)^2 
    \dfrac{\Phi\left(r\right)}{r}, \tag5.6
$$
and if $s$ is reasonable, we obtain
$$
     prob\left(T^{'}, p, q\right) >
     \dfrac{1}{poly\left(\log r\right)}. \tag5.7      
$$

To complete this half of generalizing
the two quantum Fourier tranforms over $p$,
suppose we measure $k^{'} = \lfloor\frac{qk}{p}\rfloor$.
Then 
$$
  \bigg|k^{'} - \dfrac{qk}{p}\bigg| < 1 
  \tag5.8
$$
implies 
$$ 
  \bigg|\dfrac{k^{'}}{q} - \dfrac{jt}{rt}\bigg| = 
  \bigg|\dfrac{k^{'}}{q} - \dfrac{j}{r}\bigg|
   < \dfrac{1}{q} < \dfrac{1}{2sr^2} < \dfrac{1}{2r^2}  
  \tag5.9
$$
and we can use continued fraction to find $r$
since gcd$\left(j, r\right) = 1$.

\subsubhead{\bf 5a}\endsubsubhead
We now deal with the second quantum
Fourier transform over $p$.  The technique
used here is from [3].
This generalization can be
extracted to a more general uncertainty principle
but we will not do it here because the notation
is quite cumbersome.  

We assume we have an $r^{'}$ such that 
$2r < r^{'} < 4r$.  This can be achieved
by repeating the experiment and taking 
$r^{'} = 2 + 1, 4 + 1, 8 + 1,\dots$  
Doing this will require the experiment
to be repeated at most on the
order of $\log N$ number of times,
which is polynomial.  Let 
$p^{'} = 2^l$ be such that
$$
  4r^2 < \left(r^{'}\right)^2 
   < p^{'} < 2\left(r^{'}\right)^2 <
   32r^2 ,\tag5.10
$$
and $p$ be such that
$$
  4r^2 \leq  p = rt \leq p^{'} < 
  r\left(t + 1\right) < 33r^2 .\tag5.11
$$
The choice of $p$ will allow 
continued fraction expansion to recover $r$
and the choice of $p^{'}$ will allow
the quantum Fourier transform
to be applied over a smooth number. 
Let 
$$\align
  {}&B = \left\{j\Bigg|
  b = x^j\bmod N, 0\leq j \leq p^{'} - 1\right\},
  \tag5.12 \\
  {}& a = \min\left\{c\in B\right\}. 
\endalign
$$
Since $p = rt \leq p^{'} < r\left(t + 1\right)$,
either 
$$
  B = 
  \left\{a, a + r, \dots , a + \left(t - 1\right)r\right\},
  \tag5.13
$$
or 
$$
  B = 
  \left\{a, a + r, \dots , a + tr\right\}.
  \tag5.14
$$
Instead of applying the quantum Fourier
transform over $p$ as in 3.2, we apply it
over $p^{'}$ and the algorithm becomes
$$
  \sqrt{\dfrac{1}{|B|}}
  \sum_{j\in B} |j\rangle \to
  \dfrac{1}{\sqrt{q|B|}}
  \sum_{c = 0}^{q-1}
  \sum_{j\in B} 
   \exp{\left(2\pi ijc/q\right)}|c\rangle .
   \tag5.15
$$
With 5.15, the probability of measuring
$k^{'}$ is either
$$
  prob\left(k^{'}, p^{'}, q\right) =
  \dfrac{1}{qt}\Bigg|
  \sum_{j = 0}^{t - 1}
  \exp{\left(2\pi ijr k^{'}/q\right)}
  \Bigg|^2, \tag5.16
$$
or
$$
  prob\left(k^{'}, p^{'}, q\right) =
  \dfrac{1}{q\left(t + 1\right)}\Bigg|
  \sum_{j = 0}^{t}
  \exp{\left(2\pi ijr k^{'}/q\right)}
  \Bigg|^2, \tag5.17
$$
depending on whether $B$ is of the form
5.13 or 5.14, respectively.  If we get
5.16, then nothing has changed by appling
the quantum Fourier transform over $p^{'}$
and the result will be the same as that
of 5.6, i.e.
$$
  prob\left(k^{'}, p^{'}, q\right)
  = prob\left(k^{'}, p, q\right)
   > \left(1 - \dfrac{\pi}{s}\right)^2
    \dfrac{\Phi\left(r\right)}{r}. \tag5.18
$$

We now consider the case of 5.17.  For notation
convenience, let 
$$\align
  {}& a = \sum_{j = 0}^{t - 1}
  \exp{\left(2\pi ijr k^{'}/q\right)}, \tag5.19 \\
  {}& b = \exp{\left(2\pi itr k^{'}/q\right)}. 
\endalign
$$
Notice that
$$
  \dfrac{1}{\sqrt{sr}}
  \left(1 - \dfrac{\pi}{s}\right) <
  \sqrt{prob\left(k^{'}, p, q\right)}
  = \dfrac{1}{\sqrt{qt}}|a| = 
   \sqrt{\dfrac{r}{qp}}|a| . \tag5.20
$$
We will choose $s$ so that
$$
  0 < 1 - \dfrac{1}{2\left(1 - \dfrac{\pi}{s}\right)}
    < 1 - \dfrac{1}{2t\left(1 - \dfrac{\pi}{s}\right)}
   \tag5.21
$$
is satisfied as well as 5.1.  
With our notations we have
$$\align
  {}& |a + b| = |a|\Bigg|1 + \dfrac{b}{a}\Bigg|
  \geq |a|\Bigg|1 - \dfrac{|b|}{|a|}\Bigg|
   = |a|\Bigg|1 - \dfrac{1}{|a|}\Bigg| = \tag5.22 \\
  {}& |a|\Bigg|1 - \sqrt{\dfrac{r}{qp}}
        \dfrac{1}{\sqrt{prob\left(k^{'}, p,q\right)}}
      \Bigg|.
\endalign
$$
The expressions in 5.20 and 5.21 tell us that
$$\align
 {}&1 - \sqrt{\dfrac{r}{qp}}
 \dfrac{1}{\sqrt{prob\left(k^{'}, p,q\right)}} >
  1 - \sqrt{\dfrac{r}{qp}}
 \dfrac{\sqrt{s}\sqrt{r}}
  {\left(1 - \dfrac{\pi}{s}\right) } > \tag5.23 \\
  {}& 1 - \left(\dfrac{\sqrt{s}r}
      {\sqrt{s}rt}\right)
      \dfrac{1}{\left(1 - \dfrac{\pi}{s}\right)} =
     1 - \dfrac{1}{2t}
      \dfrac{1}{\left(1 - \dfrac{\pi}{s}\right)} > 0,
\endalign
$$ 
and this implies
that 
$$
  |a + b| \geq |a|
   \left(1 - \dfrac{s}{2t\left(s - \pi\right)}
   \right) . \tag5.24
$$
For the expression in 5.17, we have
$$\align
  {}&prob\left(k^{'}, p^{'}, q\right) = 
  \dfrac{1}{q\left(t + 1\right)}
   |a + b| \geq
  \dfrac{t}{qt\left(t + 1\right)}
  |a|^2
   \left(1 - \dfrac{s}{2t\left(s - \pi\right)}
   \right)^2 = \tag5.25\\ 
  {}& \dfrac{t}{\left(t + 1\right)}
    prob\left(k^{'}, p, q\right)
   \left(1 - \dfrac{s}{2t\left(s - \pi\right)}
   \right)^2 > \\ 
  {}& \dfrac{t}{\left(t + 1\right)}
    \dfrac{1}{sr}\left(1 - \dfrac{\pi}{s}\right)^2
   \left(1 - \dfrac{s}{2t\left(s - \pi\right)}
   \right)^2 . \\ 
\endalign
$$
(Notice that 5.25 is an opportunity
for QAUP, but we will not pursue
it here.)
Since $4r^2 \leq p = rt < 32r^2$,
we have $4r \leq t < 32r$.  Using this
to bound 5.25, we obtain
$$
    prob\left(k^{'}, p^{'}, q\right) > 
     \dfrac{4}{33rs} 
    \left(1 - \dfrac{\pi}{s}\right)^2
   \left(1 - \dfrac{s}{2t\left(s - \pi\right)}
   \right)^2 . \tag5.26
$$
Finally, summing over all $k^{'}\in T^{'}$
(for either 5.16 or 5.17) yields
$$\align
  {}&prob\left(T^{'}, p^{'}, q\right) >
    \sum_{k^{'}\in T^{'}}
     \dfrac{4}{33rs} 
    \left(1 - \dfrac{\pi}{s}\right)^2
   \left(1 - \dfrac{s}{2t\left(s - \pi\right)}
   \right)^2 = \tag5.27\\
   {}&\dfrac{4\Phi\left(r\right)}{33rs} 
    \left(1 - \dfrac{\pi}{s}\right)^2
   \left(1 - \dfrac{s}{2t\left(s - \pi\right)}
   \right)^2 > \dfrac{1}{poly\left(\log r\right)}.
\endalign
$$ 
To recover $r$, we proceed as in 5.8 and 5.9.
   
\subhead{\bf 6. Motivation: Discrete Log, The Easy Case} 
\endsubhead  Mimicking section 3 of factoring, we study the easy case of 
the discrete log problem.   Given $x, g$ and $p$, 
the discrete log problem is to find the least $r$ 
such that $g^r \equiv x \bmod p$.

The quantum algorithm is as follows.  Prepare the superposition
$$
   \dfrac{1}{p-1}\sum_{a=0}^{p-2}\sum_{b = 0}^{p-2} |a,b,0\rangle .
   \tag6.1
$$
Compute in the third register $g^a x^{-b}$
$$
  \dfrac{1}{p-1}\sum_{a=0}^{p-2}\sum_{b = 0}^{p-2} |a,b, g^a x^{-b}\rangle .
   \tag6.2
$$
Measure the third register 
$y = g^k = g^a x^{-b} = g^{a - rb}$. 
The number of pairs $\left(a, b\right)$ such that
$a - rb \equiv k\bmod \left(p - 1\right)$ is $p - 1$
since there are $p - 1$ different values of $b$ to choose
from and that exhausts all the solution pairs.  Let
$$
B = \left\{\left(a, b\right)\in 
\Bbb Z_{p - 1}\times \Bbb Z_{p - 1}\Bigg| 
a - rb \equiv k\bmod \left(p - 1\right)\right\} , 
 \tag6.3
$$
then the state of the machine will be
$$
  \dfrac{1}{\sqrt{p - 1}}
  \sum_{\left(a, b\right)\in
  B} |a, b\rangle  ,
  \tag6.4
$$
where the third register is suppressed.
Now apply the quantum Fourier transform over $\Bbb Z_p$
on the first two registers and obtain
$$
   \dfrac{1}{\left(p-1\right)^{\frac{3}{2}}} 
   \sum_{c=0}^{p-2}\sum_{d = 0}^{p-2}
    \sum_{\left(a, b\right)\in 
     B}
    \exp{\left[2\pi i\left(ac + bd\right)/\left(p - 1\right)\right]}
    |c,d\rangle .
    \tag6.5
$$
Finally, measure the first and second register and obtain 
$|c, d\rangle$ with probability
$$
  \left|\dfrac{1}{\left(p-1\right)^{\frac{3}{2}}}
   \sum_{a, b, a-rb\equiv k} 
   \exp{\left[2\pi i\left(ac + bd\right)/\left(p - 1\right)\right]}
   \right|^2  .
   \tag6.6
$$
Substituting $a\equiv k + rb\bmod \left(p - 1\right)$ gives
$$
  \left|\dfrac{1}{\left(p-1\right)^{\frac{3}{2}} }
   \sum_{b=0}^{p-2}
   \exp{\left[2\pi i\left(kc + b\left(d + rc\right)\right)/\left(p - 1\right)\right]}
   \right|^2 .
   \tag6.7
$$
Notice that the sum is
$$
  \cases 0, & \text{if } d + rc \not\equiv 0\bmod \left(p-1\right) \\
   \left(p - 1\right)\exp{2\pi ikc/\left(p-1\right)}, & \text{if }
    d\equiv -rc .
\endcases
 \tag6.8
$$
Hence, with probability
$1/\left(p - 1\right)$
we will measure $c\bmod\left(p-1\right)$ and $d\equiv -rc$.
If $c$ and $p-1$ are relatively prime, we can find $r$.
Thus,  
we are interested in the set
$$
  T = \left\{\left(c, d\right)\in \Bbb Z_{p-1}\times
  \Bbb Z_{p - 1}\Bigg|
  d \equiv -rc \bmod \left(p - 1\right), gcd\left(c, p-1\right) = 1
  \right\} ,
   \tag6.9
$$
and the probability of measuring something in $T$
is 
$$
  \dfrac{\Phi\left(p-1\right)}{p - 1}
  > \dfrac{1}{\log{\log{p}} } .\tag6.10
$$

\subhead{\bf 7. QAUP - version 2, Multi-Dimensional} 
\endsubhead
Motivated by the discrete log problem, we will derive 
multi-dimensional
quantum algorithm uncertainty principles.  The quantum algorithms
that we are interested in are of the following form.  Start
with $n + 1$ registers all set to $0$:
$|0, 0, \dots , 0\rangle|0\rangle$.  For each
of the first $n$ registers,
apply the quantum Fourier
transform over $\Bbb Z_{p_{j} - 1}$ and obtain
$$
  \prod_{j = 1}^{n}
  \dfrac{1}{\sqrt{p_{j} - 1}}
  \sum_{a_{1},\dots ,a_{n}} 
   |a_{1}, \dots ,a_{n}\rangle |0\rangle .
   \tag7.1
$$
Next, compute
$$ 
  \prod_{j = 1}^{n}
  \dfrac{1}{\sqrt{p_{j} - 1}} 
  \sum_{a_{k},\dots ,a_{k}} 
   |a_{k}, \dots ,a_{k}\rangle 
   |f\left(a_{k}, \dots ,a_{k}\right)\rangle ,
   \tag7.2
$$ 
and then measure the $n + 1$ 
register.   The computer will go into the state
$$  
  \dfrac{1}{\sqrt{|B|}} 
  \sum_{|a_{1},\dots ,a_{n}\rangle\in B}  
   |a_{1}, \dots ,a_{n}\rangle   ,
   \tag7.3
$$  
where the $n + 1$ register is suppressed.
Next, zero-pad $p_j - 1$ up to $q_j$
for $q_j \geq p_j - 1$ and then apply the
quantum Fourier transform again and obtain 
$$
   \align
    \dfrac{1}{\sqrt{|B|} }
    \left(\prod_{j = 1}^{n} 
  \dfrac{1}{\sqrt{q_j} } \right)
  {}& \sum_{b_{1},\dots ,b_{n}} \Bigg\{ \tag7.4 \\
  {}&  \sum_{\left(a_{1},\dots ,a_{n}\right)\in B} 
   \prod_{l = 1}^n
   \exp\left(
       \frac{2\pi i a_{l} b_{l}}{q_{l}}
    \right)\Bigg\}
  |b_{1}, \dots ,b_{n}\rangle  .
 \endalign
$$
Finally, measure the rest of the registers and
we would like to have some sense of the probability
of measuring something in a particular set $T$.

Let $f = \otimes_{j = 1}^{n} |0\rangle$, then the
probability of measuring something in $T$ is given by
$$
  \dfrac{1}{|B|}
  ||P_T^{q} R_B^{q} f||_2^2  
  \prod_{j = 1}^{n}
  q_{j} , \tag7.5
$$
where $P_T^q$ and $R_B^q$ are the multidimensional
time and band-limiting operators respectively.  
Notice that for any $k\in T$, the probability of measuring
$k$ is given by
$$
  \dfrac{1}{|B|}\left(\prod_{j = 1}^n
    \dfrac{1}
    {q_{j} }
  \right)
  \Bigg|\sum_{\left(a_{1},\dots ,a_{m}\right)\in B}
   \prod_{l = 1}^n
   \exp\left(    
       \frac{2\pi i a_{l} k_{l}}{q_{l} }
    \right)\Bigg|^2  .
    \tag7.6
$$

\proclaim{Lemma 7.1}
Let $q_j \geq p_j - 1, 1 \leq j \leq n$,
$T\subseteq\Bbb Z_{p_1 - 1}\otimes\dots\otimes
\Bbb Z_{p_n - 1}$,
$$
  T^{'} = \Bigg\{
   k^{'} = \left(k_1^{'}, \dots , k_n^{'}\right)
         \in \otimes_{j=1}^n 
         \Bbb Z_{q_j}\Bigg| 
         k_{j}^{'} = 
         \frac{\left(q_j\right)k_j}{\left(p_j - 1\right)} +
         \epsilon_j, \left(k_1,\dots ,k_n\right)\in T 
  \Bigg\}, 
  \tag7.7
$$
$$ 
  \align
  {}& \bar p = \prod_{j=1}^n
   \left(p_{j} - 1\right), \\
  {}& \bar q = \prod_{j=1}^n
   q_{j},
 \endalign 
$$
and $k^{'}\in T^{'}$ which corresponds to $k\in T$.
Suppose 
$$
    0 \leq \dfrac{2\pi}{\bar q} 
            \sum_{|a_{1},\dots ,a_{n}\rangle 
             \in B} 
               \Bigg|\sum_{l=1}^n 
              \dfrac{a_{l}\epsilon_{l}}{q_l}
                \Bigg|
          \leq
    \dfrac{2\pi\delta_{k,B} }{\bar q} \leq 
   \dfrac{\bar p}{\bar q}||P_k^{p} R_B^{p} f||_2, 
   \tag7.8
$$
then
$$
  \align
   0 {}& \leq \dfrac{\bar p}{\bar q}||P_k^{p} R_B^{p} f||_2 -
         \dfrac{2\pi\delta_{k,B}}{\bar q} 
   \leq 
  ||P_{k^{'}}^{q} R_B^{q} f||_2. \tag7.9
\endalign
$$

\endproclaim
\demo{Proof} The proof is similar to that of lemma 4.1.   
$$\align
  {}& \dfrac{\bar p}{\bar q}||P_k^{p} R_B^{p} f||_2 - 
  ||P_{k^{'}}^{q} R_B^{q} f||_2 \leq  \tag7.10 \\ 
  {}& \dfrac{1}{\bar q}
  \Bigg|
     \sum_{|a_{1},\dots ,a_{n}\rangle\in B}
   \prod_{l = 1}^n
   \exp\left(
       \frac{2\pi i a_{l} k_{l}}{p_{l} - 1}
    \right)   -
   \prod_{l = 1}^n
   \exp\left(
       \frac{2\pi i a_{l} k_{l}^{'}}{q_{l}}
    \right) \Bigg| .
\endalign
$$
The right-hand-side of inequality 7.10 can be written as
$$
  \align
  {}&\dfrac{1}{\bar q} \Bigg|
     \sum_{|a_{1},\dots ,a_{n}\rangle\in B}
   \left[\prod_{l = 1}^n
   \exp\left(
       \frac{2\pi i a_{l} k_{l}}{p_{l} - 1}
    \right)\right]  
      \left[ \prod_{l=1}^n\exp\left(    
       \frac{2\pi i a_{l} \epsilon_{l} }{q_{l}}
    \right) - 1\right] \Bigg| .
    \tag7.11
\endalign
$$
Using
$$
    |\exp{\left(ix\right)} - 1| \leq |x|, 
    \tag7.12
$$
7.11 is bounded from above by
$$
  \align
  {}&\dfrac{1}{\bar q} 
     \sum_{|a_{1},\dots ,a_{n}\rangle\in B}
    \Bigg|
      \left[ \prod_{l=1}^n\exp\left(   
       \frac{2\pi i a_{l} \epsilon_{l} }{q_{l}}
    \right) - 1\right] \Bigg| \leq
    \tag7.13 \\
  {}&\dfrac{2\pi}{\bar q} 
     \sum_{|a_{1},\dots ,a_{n}\rangle\in B}
      \Bigg|
      \sum_{l=1}^n \dfrac{a_l\epsilon_l}{q_l}
      \Bigg| \leq 
      \dfrac{2\pi\delta_{k,B,}}{\bar q}.
\endalign
$$

The expressions in 7.10, 7.11, and 7.13 imply
$$
  \align
   {}& \dfrac{\bar p}{\bar q}||P_k^{p} R_B^{p} f||_2 -
      \dfrac{2\pi\delta_{k,B} }{\bar q}
       \leq 
  ||P_{k^{'}}^{q} R_B^{q} f||_2 . \qed \tag7.14 
\endalign
$$
\enddemo

\proclaim{Theorem 7.1 QAUP - version 2}
Under the conditions of lemma 7.1,
$$ \align
  0 {}& \leq \dfrac{\bar p}{\bar q}
   \left( 
     \dfrac{\sqrt{\bar p}}{\sqrt{|B|}}
   ||P_k^pR_B^p f||_2 -   
   \dfrac{2\pi\delta_{k,B} }
    {\sqrt{|B|\bar p} }
                    \right)^2   \leq 
      \dfrac{\bar q}{|B|}  
      ||P_{k^{'}}^qR_B^q f||_2^2 . \tag7.15
\endalign
$$
\endproclaim
\demo{Proof} Similar to theorem 4.1. \qed
\enddemo

\subhead{\bf 8. Application to Discrete Log, QAUP-version 2.} 
\endsubhead  
We will apply QAUP-version 2 to the discrete log algorithm.
Instead of applying the 
quantum Fourier transforms over $p$ as given in 6.5,
we will apply it over $q = 2^l$.  This is natural since
the dimension of the Hilbert space in qubit quantum computing
is a power of two.
We take 
$n=2$,
$p_1 - 1 = p_2 - 1 = p - 1$, $|B| = p - 1$ and 
$$     \dfrac{\sqrt{\bar p}}{\sqrt{|B|}}
   ||P_k^pR_B^p f||_2  = \dfrac{1}{\sqrt{p - 1}},
   \tag8.1
$$ 
where 8.1 comes from 6.7 and 6.8.
Let 
$$  
  q_1  = q_2  = q = 2^l = s\left(p - 1\right) > 
   \left(p - 1\right),
   \tag8.2
$$
where $s$ satisfies
$$
  0 < 1 - \dfrac{3\pi}{s} ,
  \tag8.3
$$
and let 
$$
  T^{'} = \left\{\left(
    \lfloor\dfrac{\left(q \right)k_1}{p-1}\rfloor,
    \lfloor\dfrac{\left(q \right)k_2}{p-1}\rfloor
    \right)
    \Bigg| \left(k_1, k_2\right)\in T\right\} ,
  \tag8.4
$$ 
where $T$ is given by 6.9.  With this choice of 
$T$ and $T^{'}$, we have 
$|\epsilon_1|, |\epsilon_2| < 1$.
To satisfy 7.9, we have
$$  
  \sum_{|a_0,a_1\rangle\in B}
   \Bigg|\sum_{l=0}^1 
    \dfrac{a_l\epsilon_l}{q_l}\Bigg| =
   \dfrac{1}{q}
   \sum_{a,b,a-rb\equiv k} |a\epsilon_1 + b\epsilon_2| \leq
   \dfrac{1}{q}\sum_{a,b,a-rb\equiv k} a + b .\tag8.5
$$
Since $a\equiv \left(k + rb\right)\bmod \left(p - 1\right)$,
$a$ satisfies $a < p-1$ and we can use it
to bound the last sum as follows
$$\align
   {}&\sum_{a,b,a-rb\equiv k} a + b <
  \sum_{b=0}^{p-2} p - 1 + b =
  \left(p-1\right)^2 + 
  \sum_{b=0}^{p-2} b < \tag8.6 \\
  {}&\left(p-1\right)^2 + 
   \dfrac{\left(p - 1\right)^2}{2} = 
   \dfrac{3\left(p - 1\right)^2}{2} =
   \dfrac{3\bar p}{2} .
\endalign
$$ 
This gives 
$$
  \dfrac{2\pi\delta_{k,B}}{\bar q} =
  \dfrac{3\pi\bar p}{q\bar q} =
  \dfrac{3\pi}{s^3\left(p - 1\right)}. \tag8.7 
$$
Further,
$$\align
  {}&\dfrac{\bar p}{\bar q}||P_k^{p} R_B^{p} f||_2 = 
  \dfrac{1}{s^2}
   \dfrac{\sqrt{p-1}}{p-1} 
   \dfrac{p-1}{\sqrt{p-1}}||P_k^{p} R_B^{p} f||_2 = \tag8.8\\
   {}&\dfrac{1}{s^2}
   \dfrac{\sqrt{p-1}}{p-1} 
   \dfrac{1}{\sqrt{p-1}} =
   \dfrac{1}{s^2\left(p-1\right)}.  
\endalign
$$
The expressions in 8.3, 8.7 and 8.8 imply that 
7.8 is satisfied.  

We now proceed to evaluate 7.15.  We have
$$\align
  {}&\dfrac{1}{s^2}\left(
   \dfrac{1}{\sqrt{p-1}} -
   \dfrac{2\pi}{\sqrt{\left(p-1\right)^3}}
   \dfrac{3\left(p-1\right)^2}{2s\left(p-1\right)}
   \right)^2 = \tag8.9\\
  {}&\dfrac{1}{s^2\left(p-1\right)}
   \left(1 - \dfrac{3\pi}{s}\right)^2  <
      \dfrac{\bar q}{|B|}
      ||P_{k^{'}}^qR_B^q f||_2^2 .
\endalign
$$
Finally, summing over all $k^{'}\in T^{'}$ yields
$$
  \sum_{k^{'}\in T^{'}}
      \dfrac{\bar q}{|B|}
      ||P_{k^{'}}^qR_B^q f||_2^2  > 
     \dfrac{\Phi\left(p-1\right)}
      {s^2\left(p-1\right)}
   \left(1 - \dfrac{3\pi}{s}\right)^2,  \tag8.10
$$
and this tells us that if $s$ is reasonable,
then the probability of measuring an element
in $T^{'}$ is at least the inverse
of a polynomial in $\log p$.  

To finish the algorithm, we need to recover $r$.
Since we know $q$ and $p-1$, we can check if
our measurement on the first and second
register is of the form
$$
  \left(c^{'}, d^{'}\right) = 
   \left(\lfloor\dfrac{qc}{p-1}\rfloor ,
         \lfloor\dfrac{qd}{p-1}\rfloor 
  \right), \tag8.11
$$
for some $\left(c,d\right)$ (not necessarily in $T$).
We can actually find $c$ and $d$ if it is of this form.
If $\left(c,d\right) \in T$, we can recover $r$ and
the probability of this happening
is at least inverse polynomial.

\enddocument